# Prior-knowledge-informed deep learning for lacune detection and quantification using multi-site brain MRI

Bo Li[*,1], Jeroen de Bresser[2], Wiro Niessen[1,3], Matthias van Osch[2], Wiesje M. van der Flier[4,5], Geert Jan Biessels[6], Meike W. Vernooij[1,7], Esther Bron[1], for the Heart-Brain Connection Consortium

[1] Department of Radiology and Nuclear Medicine, Erasmus MC, Rotterdam, the Netherlands
[2] Department of Radiology, Leiden University Medical Center, Leiden, the Netherlands
[3] Imaging Physics, Applied Sciences, Delft University of Technology, the Netherlands
[4] Alzheimer Center Amsterdam, Department of Neurology, Amsterdam Neuroscience, Vrije Universiteit Amsterdam, Amsterdam UMC, Amsterdam, the Netherlands
[5] Department of Epidemiology & Data Science, Vrije Universiteit Amsterdam, Amsterdam UMC, Amsterdam, the Netherlands
[6] Department of Neurology, UMC Utrecht Brain Center, University Medical Center, Utrecht, Netherlands
[7] Department of Epidemiology, Erasmus MC, Rotterdam, the Netherlands
*mail.van.boli@gmail.com

**Introduction**: Lacunes of presumed vascular origin, also referred to as lacunar infarcts, are important to assess cerebral small vessel disease[1] and cognitive diseases such as dementia[2]. However, visual rating of lacunes from imaging data is challenging, time-consuming, and is rater-dependent, owing to their small size, sparsity and mimics. Whereas recent developments in automatic algorithms have shown to make the detection of lacunes faster while preserving sensitivity, they also showed a large number of false positives[3,4], which makes them impractical for use in clinical practice or large-scale studies. Here, we develop a novel framework that, in addition to lacune detection, outputs a categorical burden score. This score could provide a more practical estimate of lacune presence that simplifies and effectively accelerates the imaging assessment of lacunes. We hypothesize that the combination of detection and the categorical score makes the procedure less sensitive to noisy labels.

**Methods**: We included 3D T1w scans, FLAIR scans and manual annotations of participants having lacunes but no other infarcts from the Heart-Brain Study[5], Trace-VCI Study[6], and Valdo challenge[7] (n=122). In addition, to evaluate on negative cases, we included 18 scans without lacunes from Valdo. Based on the assessment criteria of lacunes[10], the framework outputs a dense prediction as well as a categorical burden score ('0', '1-3', or '>3' lacunes), jointly providing a full quantification of lacune presence, burden, and location (Fig. 1). The method consists of three main stages, data preprocessing, detection, and lacune quantification. Data preprocessing includes standardization of the datasets (e.g., $1mm^3$ resolution, RAS orientation), bias field correction, and scan-wise intensity normalization. To detect at high sensitivity, we propose to use a false-negative-weighted binary cross-entropy (FNw-BCE) loss function to increase the loss due to false negative predictions. Also, we propose to use the difference map between CSF-normalized T1w and FLAIR images as additional input to incorporate contrast between sequences. For the refinement and quantification of the detected lacunes, local and global supervision is obtained by a location prior and by a joint loss function that measures the agreement of the predicted lacune segmentation and burden score with those computed from the manual annotation (Fig. 1). The second and third stages are independent 5-layer U-Net[8] neural networks using data augmentation (rotation, flipping, and contrast changes). The detection sensitivity was instance (lacune)-wise evaluated with the manual annotations using stratified 10-fold cross validation. For one fold (22 test scans), we additionally evaluated the final lesion burden prediction performance using balanced classification accuracy (BCA)[9] that normalizes group size.

**Results**: The average instance sensitivity using the proposed FNw-BCE loss function (0.91 ± 0.26) was significantly higher than that using the voxel-ratio-weighted BCE loss function (0.68 ± 0.26) in 10-fold cross-validation (paired t-test, $p=5.61e^{-10}$). The mean sensitivity in each fold (solid line, Fig. 2a) was generally higher than 0.8. The BCA was 0.80 ± 0.03 over three categories ('0': 0.81, '1-3': 0.82, and '>3': 0.75). The confusion matrix (Fig. 2b) shows that the majority of the scans (17/22) were accurately classified, showing that the quantified burden score is relatively robust to the few FPs (Fig. 1a). The dense prediction was overall plausible compared with the manual annotations.

**Conclusions**: We developed and evaluated a new method for fully automatic detection and quantification of lacunes on a multi-site dataset. The method has the potential of effectively accelerating and simplifying the assessment of cerebral lacunes from imaging data. Some misclassification can be contributed to the visual rating, which is highly challenging, as demonstrated by the example in Fig. 2c. A better evaluation of those uncertain cases will be the focus of our future work.

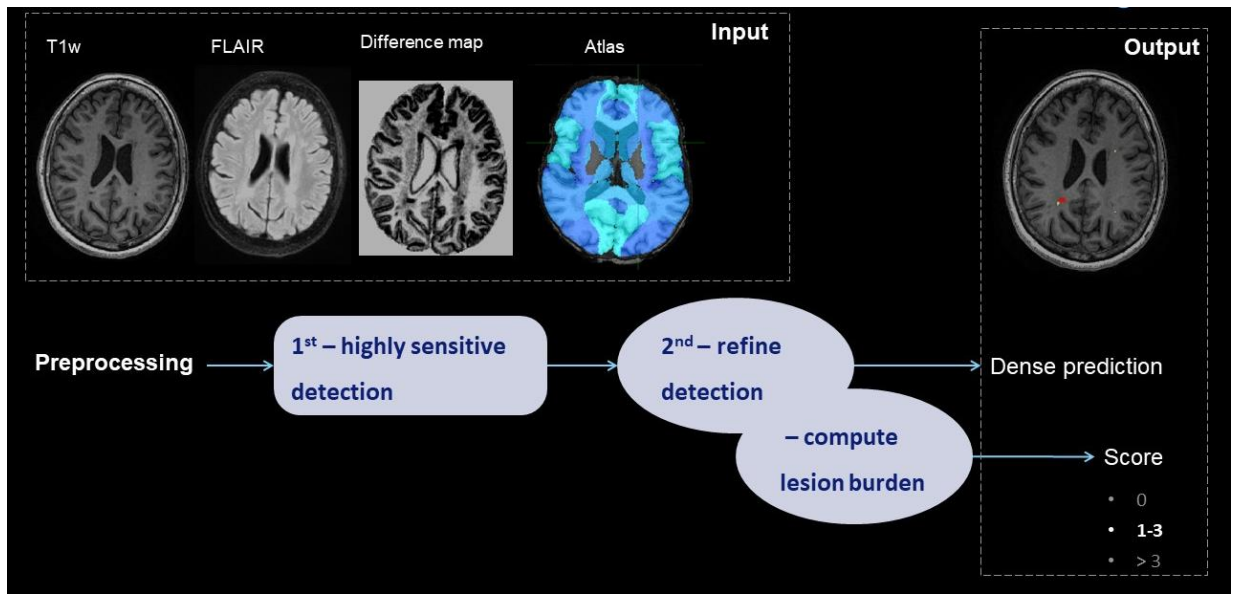

Fig. 1. Diagram of the proposed framework. For location information, we use the Hammers atlas (Hammers et al., Hum Brain Mapp 2003). For better visualization, the output dense prediction (yellow) and manual annotation (red) are overlaid on the T1w image.

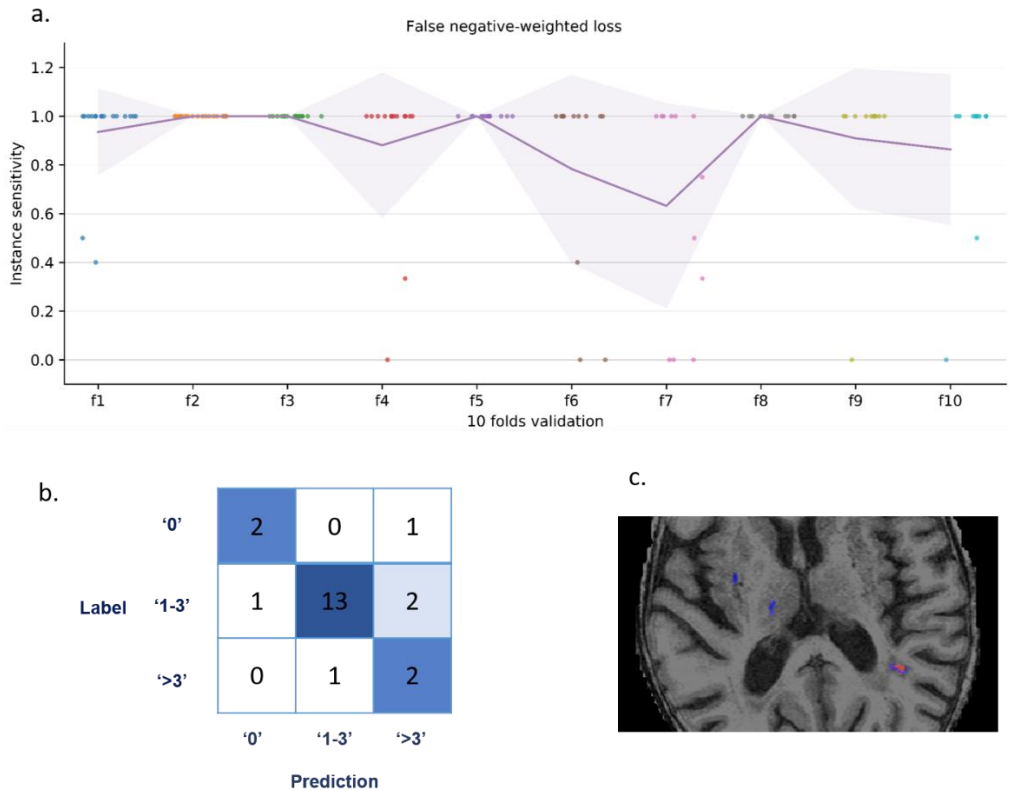

Fig 2. Performance of the proposed lacune detection algorithms using the false-negative-weighted BCE loss function (a) instance sensitivity over 10 folds; (b) confusion matrix of the test data; and (c) rater disagreement on an example ROI (Rater A (blue): 11 lacunes in total, rater B (red): 2 lacunes in total).

**Reference** (max 10)